\documentclass[preprint,showpacs,preprintnumbers,pra,floatfix,superscriptaddress,amsmath,amssymb]{revtex4}
\usepackage{graphicx}
\usepackage[latin1]{inputenc}

\newcommand{\dept}{\left(t\right)}
\newcommand{\depr}{\left(\mathbf{r}\right)}

\newcommand{\deprt}{\left(\mathbf{r},t\right)}

\newcommand{\be}{ \begin{equation} }
\newcommand{\ee}{ \end{equation} }
\newcommand{\bea}{\begin{eqnarray} }
\newcommand{\eea}{\end{eqnarray} }
\newcommand{\br}{\mathbf{r}}
\begin{document}
\title{Order parameter for the dynamical phase transition in Bose-Einstein
condensates with topological modes}
\author{E. R. F. Ramos}\email{edmir@ursa.ifsc.usp.br}
\affiliation{Instituto de Física de São Carlos, Universidade de São
Paulo, Caixa Postal 369, 13560-970, São Carlos-SP Brazil}
\author{L. Sanz}\altaffiliation{Current adress at Instituto de Física, Universidade
Federal de Uberlândia, Caixa Postal 593, 38400-902, Uberlândia-MG,
Brazil}\affiliation{Instituto de Física de São Carlos, Universidade
de São Paulo, Caixa Postal 369, 13560-970, São Carlos-SP Brazil}
\author{V.I. Yukalov}
\affiliation{Bogolubov Laboratory of Theoretical Physics, Joint
Institute for Nuclear Research Dubna 141980, Russia}
\author{V. S. Bagnato}
\affiliation{Instituto de Física de São Carlos, Universidade de São
Paulo, Caixa Postal 369, 13560-970, São Carlos-SP Brazil}
\begin{abstract}

In a trapped Bose-Einstein condensate, subject to the action of an
alternating external field, coherent topological modes can be
resonantly excited. Depending on the amplitude of the external field
and detuning parameter, there are two principally different regimes
of  motion, with mode locking and without it. The change of the
dynamic regime corresponds to a dynamic phase transition. This
transition can be characterized by an effective order parameter
defined as the difference between fractional mode populations
averaged over the temporal period of oscillations. The behavior of
this order parameter, as a function of  detuning, pumping amplitude,
and atomic interactions is carefully analyzed. A special attention
is payed to numerical calculations for the realistic case of a
quadrupole  exciting field and the system parameters accessible in
current experiments.

\end{abstract}

 \maketitle

\section{Introduction}
\label{sec:intro}

Bose-Einstein condensation (BEC) in weakly interacting dilute Bose
gases is a topic of high current interest and intensive
investigations, both experimental and theoretical, as can be
inferred form the book \cite{Pitaevskii03} and review articles
\cite{{Dalfovo99},{Courteille01},{Yukalov04},{Andersen04},{Bongs04},{Yukalov05},{Posazhennikova06}}.
At sufficiently low temperature, the weakly interactiong gas is
almost completely Bose-condensed and is well described by the
Gross-Pitaevskii equation \cite{Pitaevskii03}. Then the system is in
its ground state. One could pose a question whether nonground-state
Bose-Einstein condensates could be realized?

The possibility of creating nonground-state Bose-Einstein
condensates was advanced in Ref. \cite{Yukalov97}. As is clear, such
a condensate has to be nonequilibrium, since in an equilibrium
system at zero temeprature, the state with the lowest energy would
be always preferable. The most important point is that the atomic
system would possess a discrete spectrum, or at least, would have
finite gaps between different parts of its spectrum, such as occur
in optical lattices. A straightforward way of obtaining a purely
discrete spectrum is by confining atoms in a trapping potential.
Then, in order to transfer atoms  from their ground BEC state to
another nonground level, the system has to be subject to an
oscillatory external field. The field oscillation is to be close to
the frequency of the transition between the ground and the chosen
excited state of the trap. The states of the trapped coherent atomic
cloud are given by the set of solutions to the stationary
Gross-Pitaevskii equation, which are termed topological coherent
modes. Due to  the nonlinearity of the system, caused by atomic
interactions, the spectrum is not equidistant and allows for the
selection of the desired levels by means of the resonant modulation
of an external field. It has been mathematically accurately proved
\cite{{Yukalov97},{Yukalov02}} that this resonantly excited
condensate can be well approximated  by an effective two-level
system. The analytical proof has been confirmed
\cite{Yukalov-Marzlin04} by direct numerical simulations for the
time-dependent Gross-Pitaevskii equation. The topological coherent
modes have also been studied theoretically in Refs.
\cite{{Ostrovskaya00},{Kivshar01},{D'Agosta02},{D'Agosta-Presilla02},{Proukakis02},{Adhikari04}}.
The  generation of a nonground-state BEC has recently been observed
in optical lattices \cite{Muller07}.

The resonant system with nonground-state BEC, generated by an
applied modulating field, possesses several unusual properties, as
is reviewed in Ref. \cite{Yukalov06}. One of the interesting
features is the existence of two essentially different regimes of
motion. In one of them, the fractional mode populations are locked
in the regions close to their initial conditions, while in another
regime, the mode populations oscillate in the whole diapason between
zero and one. The transition between these two dynamical regimes
represents a kind of a dynamic phase transition. It has been shown
\cite{{Yukalov00},{Yukalov01},{Yukalov-Yukalova02}} that this
dynamical transition is equivalent to a phase  transition in an
effective averaged system, accompanied by critical phenomena around
the phase transition. According to Landau and Lifshitz
\cite{Landau80}, an order parameter is a quantity defined in such a
way that it takes non-zero values, positive or negative, in the
unsymmetrical phase and is zero in the symmetrical phase. So,
similarly to phase transitions in equilibrium systems, it is
possible to define an order parameter characterizing the dynamic
phase transition. The role of such an order parameter for the system
with resonantly generated topological modes is played by the
difference between the time-averaged mode populations for the ground
and the excited states.

The aim of the present work is twofold. First, we analyze the
order-parameter behavior for a wide range of different values of the
detuning, pumping, and interaction strengths. Second, we accomplish
a numerical investigation of the order-parameter features for a
setup typical for current experiments, so that to choose realistic
conditions for experimentally realizing the discussed generation of
the coherent topological modes and observing the dynamic phase
transition between the mode-locked and unlocked regimes.

The structure of the paper is as follows. In Section \ref{sec:two},
we recall the main equations describing the resonant excitation of
the coherent topological modes \cite{{Yukalov97},{Yukalov01}},
characterizing the coupling between the ground and excited states of
BEC. In Section \ref{sec:order}, we present the results of our
investigation  for the behavior of the order parameter under the
varying amplitude of the excitation  field, detuning from resonance,
and for different strengths of atomic interactions. In section
\ref{sec:quad}, we explore the case of a realistic alternating field
that could be realized in experiment, for which we consider an
oscillatory magnetic quadrupole field. We summarize our results in
the concluding section .

\section{Main Equations and Definition of Model}\
\label{sec:two}

The mathematical model, rather accurately describing a trapped BEC
of a dilute Bose gas, is the well-known Gross-Pitaevskii
equation~\cite{Dalfovo99}, which is equivalent to the nonlinear
Schr\"odinger equation
  \be
  \hat{H}[\Phi]\Phi=i\hslash\frac{\partial\Phi}{\partial t}, \qquad
  \hat{H}[\Phi]=-\frac{\hslash^2}{2m_0}\nabla^2 + U_{trap}\depr + A_s
  \left|\Phi\right|^2,
  \label{eq:gpe}
  \ee
where $U_{trap}\depr$ is the trapping potential,
$A_s=4\pi(N-1)\hslash^2a_s/m_0$, with $m_0$  being atomic mass;
$a_s$, the s-wave scattering length; and $N$, the number of atoms.
The stationary wave-function $\Phi$ can be written in the form
  \be
  \Phi_n=\Phi_n\deprt=\varphi\depr e^{-iE_nt/\hslash},
  \label{eq:phin}
  \ee
where $\varphi_n\depr$ is a solution to the stationary
Gross-Pitaevskii equation
  \be
  \hat{H}[\varphi_n]\varphi_n=E_n\varphi_n.
  \label{eq:gpe stationary}
  \ee

Since the Hamiltonian in equation (\ref{eq:gpe stationary}) is
non-linear, the spectrum  $E_n$ is not equally
spaced~\cite{Yukalov97}. Because of this, it is possible to couple
the ground BEC state in the confining potential to another mode
(labeled $p$) by applying  an oscillatory field~\cite{Yukalov97},
e.g., given by
  \be
  V_p\deprt=V\depr cos(\omega t),
  \label{eq:vrt}
  \ee
where $\omega$ is the angular frequency characterizing the time
variation of the field  and $V\depr$ is a spatially dependent
amplitude. If this frequency is chosen to be close to the transition
frequency between the two modes, then, as has been shown
\cite{{Yukalov97},{Yukalov02}}, the total wave-function $\Phi$ can
be written as the sum of two modes
  \be
  \Phi\deprt=c_0\dept\varphi_0\depr e^{-iE_0t/\hslash}+c_p\dept\varphi_p\depr
  e^{-iE_pt/\hslash},
  \label{eq:phi2}
  \ee
where the population $n_j$ of each mode is given by
  \be
  n_j=|c_j(t)|^2.
  \label{eq:n}
  \ee
This implies that if we wish to know the temporal evolution of these
mode populations, we  need to find out how the coefficients $c_j$
evolve in time. Substituting Eq. (\ref{eq:phi2}) into the
Schr\"odinger equation
  \be
  \left[\hat{H}+V\deprt\right]\Phi\deprt=
  i\hslash\frac{\partial\Phi\deprt}{\partial t},
  \label{eq:hplusv}
  \ee
we obtain the rate equations for the coefficients $c_0$ and $c_p$,
\begin{subequations}
  \bea
  \frac{dc_0}{dt}&=&-i \alpha_{0p} n_p c_0 - \frac{i}{2}\beta
  e^{i\Delta\omega t}c_p,\\
  \nonumber\\
  \frac{dc_p}{dt}&=&-i \alpha_{p0} n_0 c_p - \frac{i}{2}\beta^*
  e^{-i\Delta\omega t}c_0.
  \eea
  \label{eq:system}
\end{subequations}

An accurate derivation of these rate equations can be done
\cite{{Yukalov97},{Courteille01},{Yukalov02}} by employing the
Krylov-Bogolubov averaging technique \cite{Bogolubov61}. In Eq.
(\ref{eq:system}), we have the transition amplitude, defined as
  \be
  \alpha_{jk}=\frac{A_s}{\hslash}\int |\varphi_j\depr|^2
  \left(2|\varphi_k\depr|^2-|\varphi_j\depr|^2\right)d\br,
  \label{eq:alpha}
  \ee
the coupling amplitude caused by the external field, defined as
  \be
  \beta=\frac{1}{\hslash}\int \varphi_0^*\depr V\depr \varphi_p\depr
  d\br,
  \label{eq:beta}
  \ee
and the detuning $\Delta\omega = \omega - \omega_{p,0}$, which is
the difference between  the field frequency and the transition
frequency, given by $\hslash\,\omega_{p,0}=E_{p}-E_{0}$. Note that
amplitude (\ref{eq:alpha}) is unsymmetric, i.e., not necessarily
$\alpha_{0p}=\alpha_{p0}$. It comes from the non-linearity
introduced by interaction. The interaction will have different
values depending on the occupied state and that may be the cause of
the obtained unsymmetry in $\alpha_{jk}$.

Our physical system is similar to a condensate in a double-well
potential. Here, the coupling field characterized by the parameter
$\beta$ has the role of tunneling and there is the same nonlinear
term associated with collisions. Instead of using the two-mode model
\cite{Milburn97}, we prefer to use the Gross-Pitaevskii formalism
which predict the form of the spatial distributions of wave function
for both, ground and excited, condensates.

\begin{figure}[ht]
\centering
\includegraphics[scale=1.5]{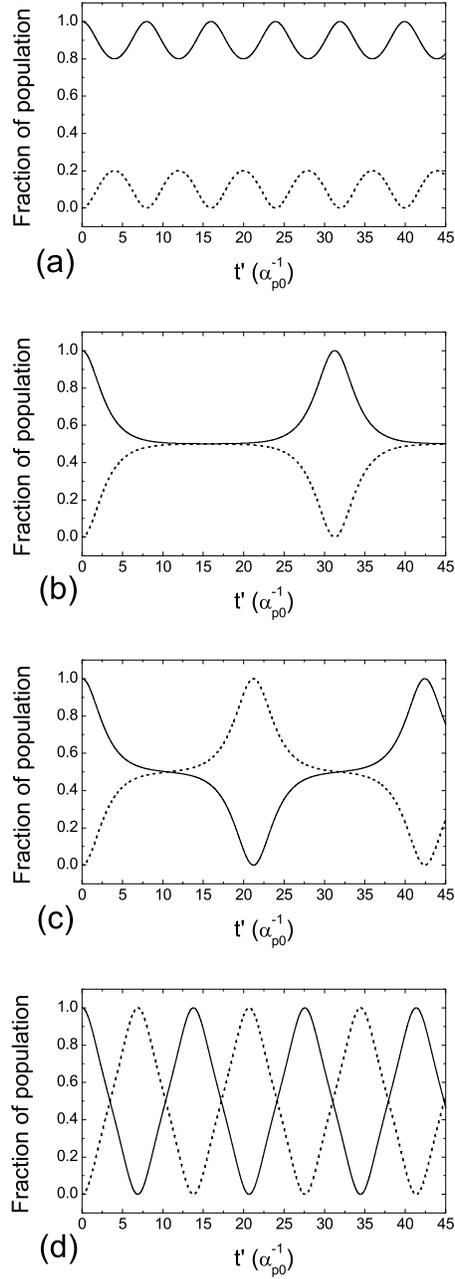}\\
\caption{Time evolution of the population fraction of the ground
state (solid line) and the excited state (dashed line) in the
presence of a resonant field, $\delta=0$, and $a=1$ for (a) $b=0.4$;
(b) $b=0.5$; (c) $b=0.5001$; (d) $b=0.6$.} \label{fig:popa1}
\end{figure}

The system of equations (\ref{eq:system}) can be solved numerically
for each given set of  parameters. In order to simplify the
description, the following dimensionless parameters  are introduced:
  \be
  a = \frac{\alpha_{0p}}{\alpha_{p0}}, \qquad
  b=\frac{\beta}{\alpha_{p0}}, \qquad
  \delta=\frac{\Delta\omega}{\alpha_{p0}},
  \ee
with the time scaling
  \be
  t'=\alpha_{p0}\,t.
  \ee
This system of equations allows us to find the temporal behavior of
the mode amplitudes characterizing the fractional mode populations
(\ref{eq:n}) \cite{{Yukalov97},{Yukalov00}}. In  Fig.
\ref{fig:popa1}, we present the temporal evolution of the mode
populations for a resonant field ($\delta=0$), the interaction
amplitude $a=1$, and different values of the pumping amplitude $b$.
We assume that, initially, all atoms are in the ground-state BEC, i.
e., $c_0(0)=1$ and $c_p(0)=0$.

\begin{figure}[t]
\centering
\includegraphics[scale=1.5]{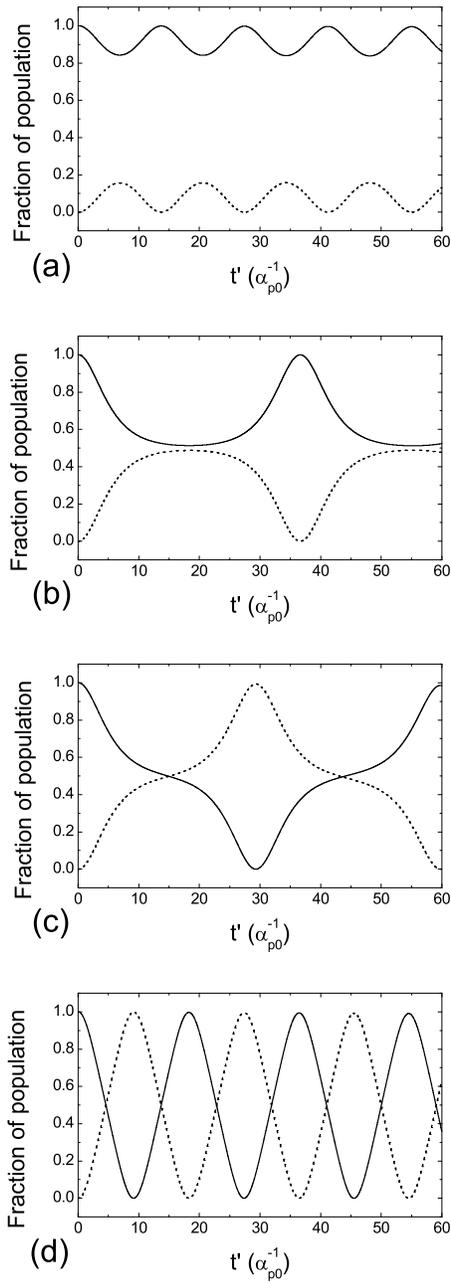}\\
\caption{Time evolution of the population fraction in the ground
state (solid line) and the excited state (dashed line) in the
presence of a field with the detuning $\delta=0.45$ and $a=0.1$ for
(a) $b=0.2$; (b) $b=0.2749$; (c) $b=0.2750$; (d) $b=0.4$.}
\label{fig:popa01}
\end{figure}

We see that there are two different types of behavior of the
population dynamics. For  $b<0.5$, the ground state population is
always larger than that of the excited state. While for $b>0.5$, the
time averages of both population modes are the same, which means
that it is possible to transfer all atoms from the ground to the
excited state. The same effect can be obtained for other values of
$a$ as we can see in Fig. \ref{fig:popa01},  with $a=0.1$ and a
detuning $\delta=0.45$.

As is seen, it is easier to detect a trapped BEC in an excited
state, when the dynamics   correspond to the mode-unlocked regime
shown in the cases (c) and (d) of Figs. \ref{fig:popa1} and
\ref{fig:popa01}, since in these cases the system stays longer in
the  excited state. In order to deeper understand how the excitation
depends on the parameters of the system, it is useful to analyze the
behavior of the order parameter, as defined in Refs.
\cite{{Yukalov01},{Yukalov-Yukalova02}}.

\section{Order Parameter for Dynamic Transition}
\label{sec:order}

The order parameter, characterizing the dynamic phase transition for
the case of the resonant BEC, with the generated topological modes,
can be defined \cite{Yukalov01}, as the difference $\eta$ between
the time-averaged fractional mode populations,
  \be
  \eta=\overline{n}_0-\overline{n}_p,
  \label{eq:eta}
  \ee
where $\overline{n}_i$ is the time average of $n_i$ over the whole
oscillation period.  This order parameter clearly shows us what kind
of dynamic regime occurs. Observing the character of the mode
oscillations, we note that in cases (a) and (b) of Fig.
\ref{fig:popa1} and Fig. \ref{fig:popa01}, $\eta>0$, and in the (c)
and (d) cases, $\eta=0$.

\begin{figure}[ht]
\includegraphics[width=0.5\textwidth]{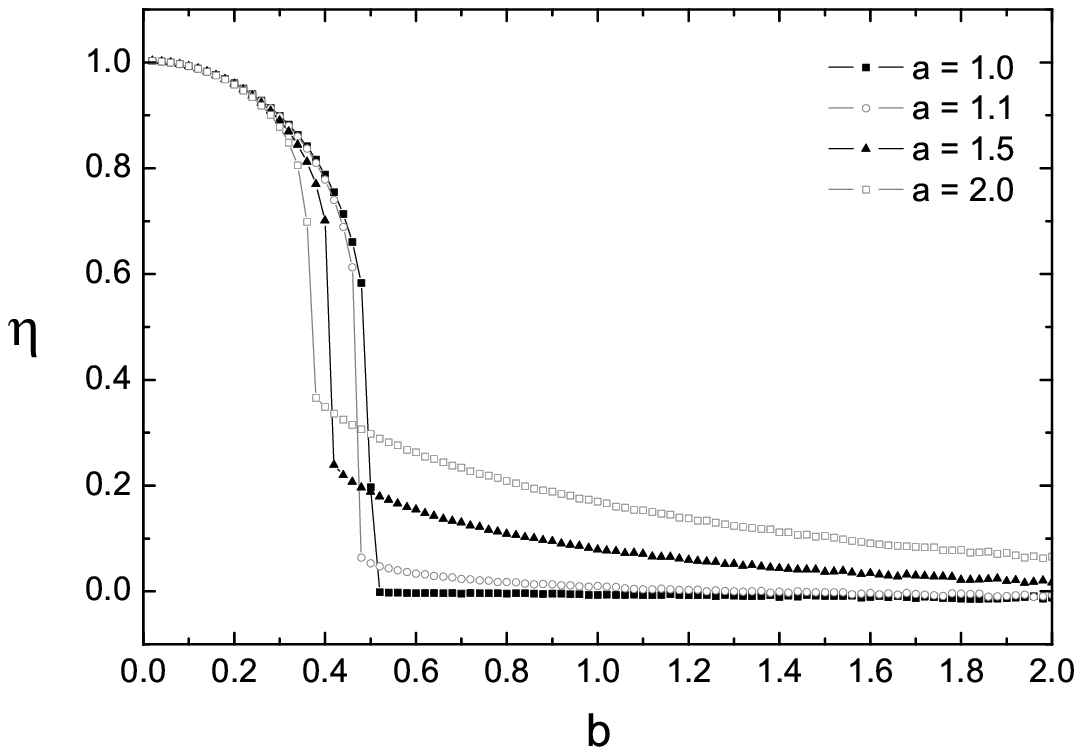}
\includegraphics[width=0.5\textwidth]{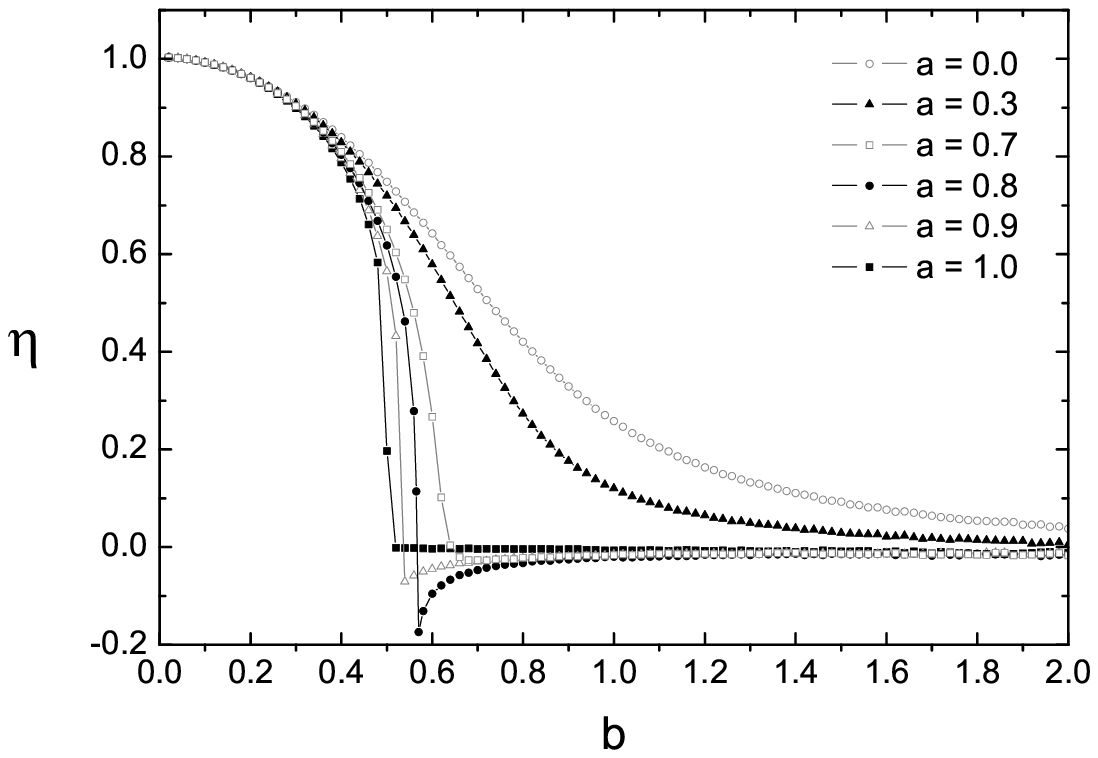}
\caption{Order parameter $\eta$ as a functions of $b$ for different
values of $a$ with a  resonant field excitation.} \label{fig:etaa}
\end{figure}

We have calculated the behavior of $\eta$ as a function of $b$, for
$0\leq a \leq 2$ with  different amplitudes of the resonant field.
These results are shown  in Fig. \ref{fig:etaa}, from where we see
that there is an abrupt change of the order  parameter for
$a\geq0.8$, while for $a\leq0.8$, $\eta$ goes smoothly to zero.
Also, for $0.8\leq a< 1$, $\eta$ becomes negative for some values of
$b$. When the order parameter is negative, this does not mean that
the fraction of the excited mode is always larger than the ground
state population. It simply shows that the BEC on the average spends
more time in the excited  state, which would favor the detection of
condensed atoms in that state.

\begin{figure}[t]
\centering
\includegraphics[scale=1.5]{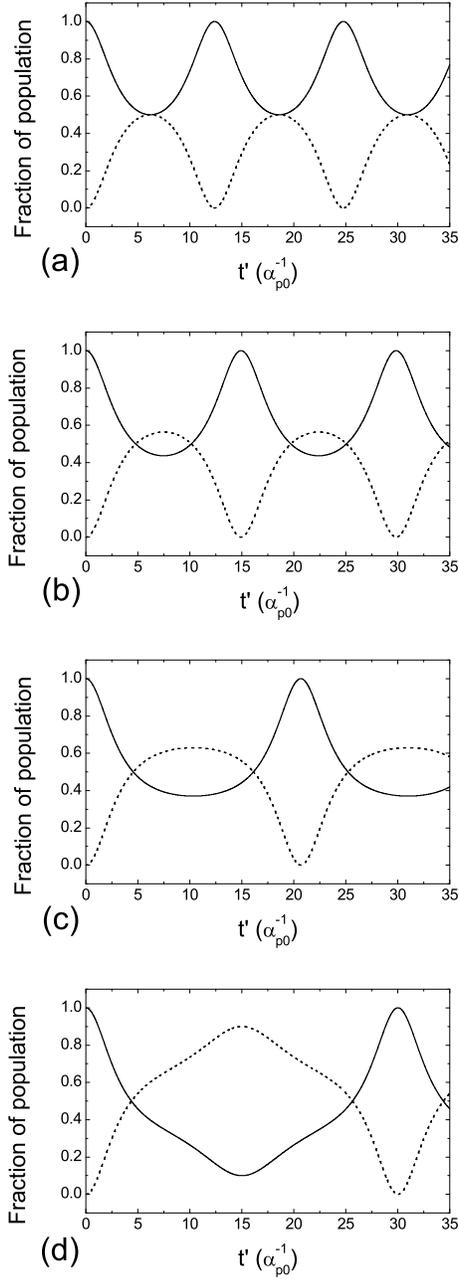}\\
\caption{Fractions of mode populations in the ground state (solid
line) and the excited state (dashed line) in the presence of a
resonant field, $\delta=0$, and $a=0.8$ for (a) $b=0.55$; (b)
$b=0.56$; (c) $b=0.565$; (d) $b=0.57$.} \label{fig:popa08}
\end{figure}

To better illustrate the peculiarities of the population evolution,
we show in Fig. \ref{fig:popa08} the mode populations $n_0$ and
$n_p$, for $a=0.8$, for different values of $b$. In all these cases
we set the resonant external field ($\delta=0$). As presented in
Fig. \ref{fig:etaa} and Fig. \ref{fig:popa08}, when $a=0.8$ and
$b=0.57$, we have $\eta\simeq-0.174$. Then $n_p$ never reaches one,
i. e., we do not get a completely  excited BEC. For conditions
presented in Fig. \ref{fig:popa08}(d), the excited state population
is dominant for the major part of time.

\begin{figure}[t]
\includegraphics[width=0.5\textwidth=1]{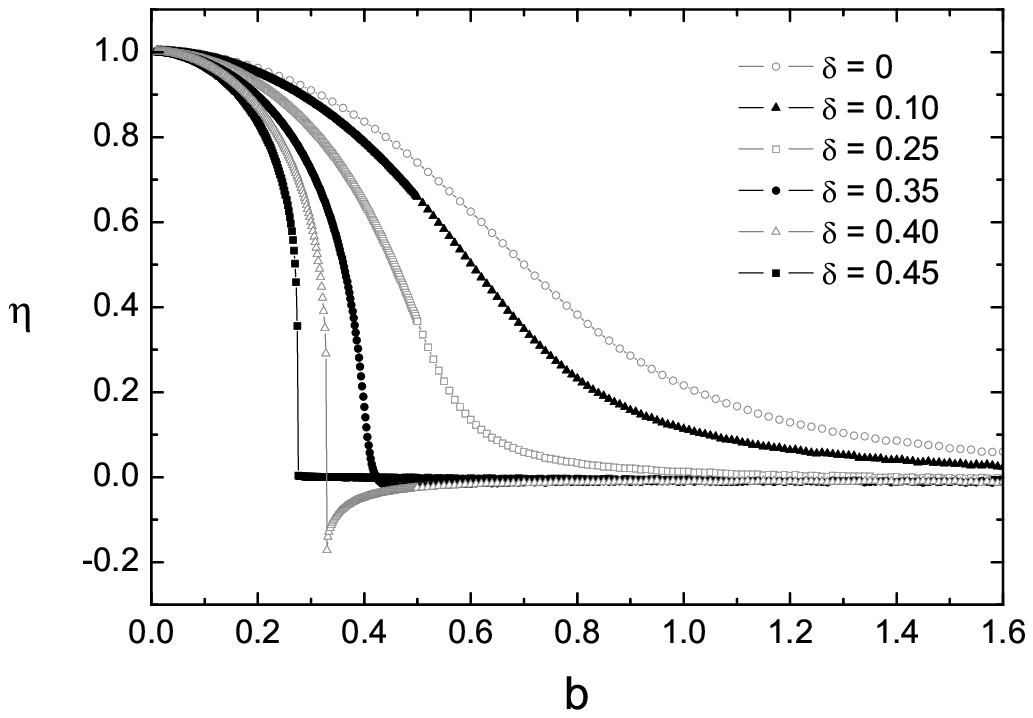}
\includegraphics[width=0.5\textwidth=1]{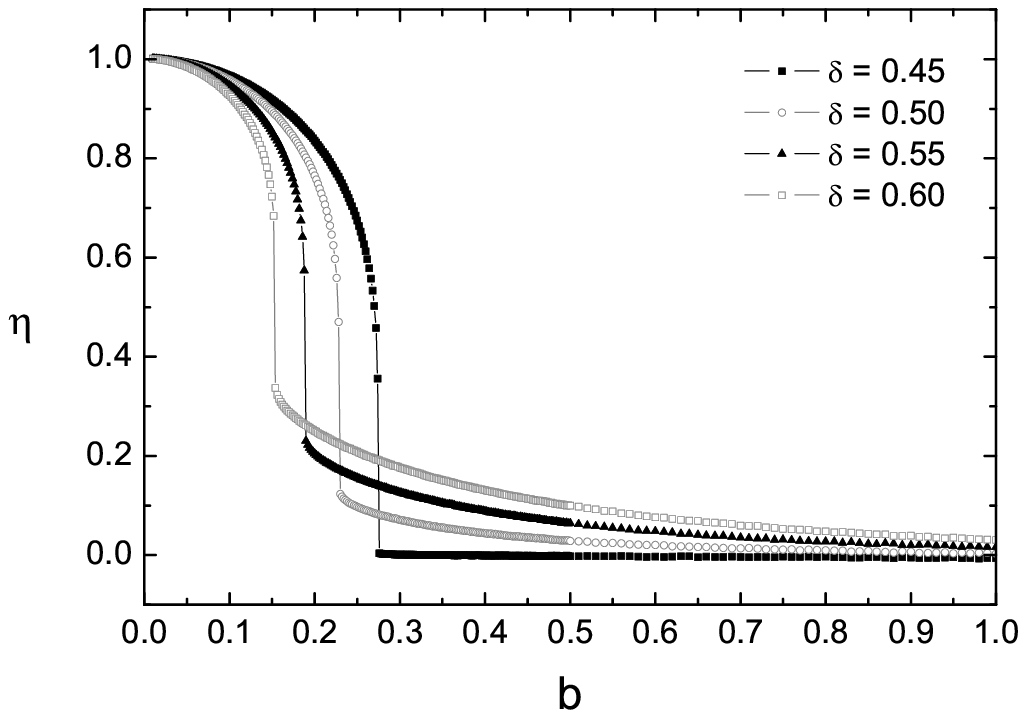}
\caption{Order parameter $\eta$ as a function of $b$ for different
values of $\delta$ with $a=0.1$.} \label{fig:etad}
\end{figure}

The behavior of the order parameter when varying the detuning for a
fixed interaction strength  $a=0.1$ is presented in Fig.
\ref{fig:etad}, which shows $\eta$ as a function  of $b$ for
different values of the detuning. This figure, together with Fig.
\ref{fig:etaa}, characterizes the overall behavior of the order
parameter under varying the system parameters $a$, $b$ and $\delta$.

The question that is yet left open is how and under what particular
conditions the  proposal of obtaining the dynamical phase transition
for a trapped BEC \cite{Yukalov01}  could be realized in laboratory.
To this end, we need to study the behavior of the order parameter
$\eta$ for a realistic setup of the applied field geometry, field
amplitude,  and detuning, so that to identify the critical field
that changes the dynamic regime of the mode excitation. The
specifications for the dimensionless parameters $a$, $b$, and
$\delta$ for a typical laboratory setup is, therefore, necessary. In
the following section, we calculate these parameters for a magnetic
quadrupole field, which corresponds to the experiment in progress in
our group, employing a $^{87}Rb$ condensate.

\section{Magnetic Quadrupole as an Excitation Field}\
\label{sec:quad}

The amplitudes $\alpha_{0p}$, $\alpha_{p0}$, and $\beta$ cannot be
obtained as exact  analytic expressions, since there are no exact
analytic solutions for the Gross-Pitaevskii equation. We can,
however, obtain rather accurate approximate solutions using the
\emph{Optimized Perturbation Theory} \cite{Yukalov76}, whose
detailed description can be found, e.g., in Ref.
\cite{Courteille01}. In this case, approximate expressions for the
ground state and three excited-state wave functions, for a harmonic
trap
 \be
 U_{trap}=\frac{1}{2}\left(\omega_r^2\,r^2+\omega_z^2z^2\right),
 \ee
are given by
  \bea
  \phi_{000}&=&\left( \frac{u_{000}^2v_{000}}{\pi^3l_r^6} \right
  )^{1/4}e^{-(u_{000}r^2+v_{000}z^2)/2l_r^2},
  \label{eq:phi000}\\
  \nonumber\\
  \displaystyle\phi_{010}&=&u_{010}\left( \frac{v_{010}}{\pi^3l_r^6}
  \right )^{1/4}\frac{r}{l_r} e^{i\varphi}
  e^{-(u_{010}r^2+v_{010}z^2)/2l_r^2},
  \label{eq:phi010}\\
  \nonumber\\
  \displaystyle\phi_{001}&=&\left( \frac{4u_{001}^2v_{001}^2}{\pi^3l_r^6}
  \right )^{1/4}\frac{z}{l_r} e^{-(u_{001}r^2+v_{001}z^2)/2l_r^2},
  \label{eq:phi001}\\
  \nonumber\\
  \phi_{100}&=&\left( \frac{u_{100}^2v_{100}}{\pi^3l_r^6} \right
  )^{1/4}\left(1-u_{100}\frac{r^2}{l_r^2}\right)
  e^{-(u_{100}r^2+v_{100}z^2)/2l_r^2}, \nonumber \\
  \label{eq:phi100}
  \eea
where $l_r=\sqrt{\hslash/m_0\omega_r}$ is the oscillator length, and
$u_{nmk}$ and $v_{nmk}$ are the variational parameters, defined by
minimizing the corresponding energy of each mode.

The alternating magnetic field in a magnetic trap can be created as
an oscillatory  quadrupole magnetic field of the form
  \be
  \textbf{B}(\textbf{r},t)=\left(Ax\hat{x}+Ay\hat{y}-
  2Az\hat{z}\right)cos\omega t,
  \label{eq:B}
  \ee
which corresponds to the field formed by a pair of coils operating
in the anti-Helmholtz configuration. Then the potential
$V(\textbf{r})$ in Eq. (4) is given by
  \be
V(\textbf{r})=-\vec{\mu}\cdot\textbf{B}=g_F\,m_F\,\mu_B\,A\,
\sqrt{r^2+4z^2},
  \label{eq:v}
  \ee
where $\vec{\mu}$ is the atomic magnetic moment, $g_F$ is the
gyromagnetic factor, $m_F$ is the magnetic quantum number, and
$\mu_B$ is the Bohr's magneton. It is easy to see that the sole
topological mode, from the lowest modes given above, which could be
coupled to the ground-state mode by the potential (\ref{eq:v}), is
the radial dipole mode $\phi_{100}$. For this mode, we have made
calculations of the order parameter $\eta$ as a function of the
field-amplitude parameter $A$. We have chosen this parameter $A$
because it is easily controllable in laboratory, since it is
proportional to the current in the coils constructed to produce the
desired field \cite{Bergeman87}.

\begin{figure}[ht]
\centering
\includegraphics[width=0.5\textwidth=1]{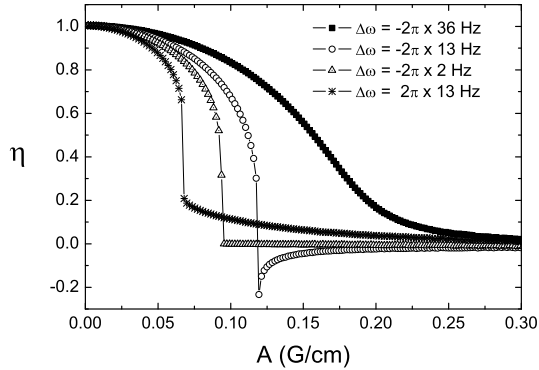}\\
\caption{Order parameter $\eta$ for the exciting magnetic quadrupole
field as a function of the field parameter $A$ for different values
of the field detuning.} \label{fig:etaquad}
\end{figure}

We set all parameters of the system according to the experiment in
progress, where $^{87}Rb$ is employed, having mass
$m_0=0.144\times10^{-21}\,g$ and the scattering length
$a_s\simeq6nm$. The magnetic trap has $\omega_r=2\pi\times120\,Hz$
and $\omega_z=2\pi\times24\,Hz$. The number of atoms is taken as
$N=10^4$, which corresponds to a realistic experimental number.
Figure \ref{fig:etaquad} shows the obtained results for different
detunings. In this case, the transition frequency between the ground
state and the radial dipole mode is
$\omega_{100,0}=2\pi\times190\,Hz$.
This figure clearly shows that there exist critical points in the
behavior of $\eta$,  which depend on the detuning. In Fig.
\ref{fig:etaquad}, when, for instance,
$\Delta\omega=-2\pi\times2\,Hz$, the critical point happens for
$A=0.096\,G/cm$. Three different situations are seen in Fig.
\ref{fig:etaquad}. For $\Delta\omega=-2\pi\times13\,Hz$, we have the
situation similar to that for $a=0.1$ and $\delta=0.40$, shown in
Fig. \ref{fig:etad}, when $\eta<0$. In this case, the critical point
occurs when $A\simeq0.119\,G/cm$. We also have the case when the
critical point occurs for $\eta>0$, which is analogous to the
behavior obtained for $a=0.1$ and $\delta>0.45$ and shown in Fig.
\ref{fig:etad}. Here, for $\Delta\omega=2\pi\times13\,Hz$, the
critical point happens for $A\simeq0.068\,G/cm$. In the case of a
negative detuning, $\Delta\omega=-2\pi\times36\,Hz$, there is no
abrupt change in the behavior of $\eta$, similarly to the case shown
in Fig. \ref{fig:etad}, where a resonant field is considered.

\section{Conclusion}\
\label{sec:conclusion}

The dynamics of the resonant excitation of topological coherent
modes of a Bose-Einstein  condensate in a harmonic trap is analyzed.
The behavior of the effective order parameter is investigated as a
function of the external field amplitude for a variety of different
system characteristics, such as the detuning from the resonance and
interaction amplitudes. An important observation is that the
detuning can compensate the influence of  atomic interactions,
producing in the case of a weak interaction ($a=0.1$), the behavior
similar to that obtained for a strong interaction ($a=1$) and a
purely resonant field. The pumping field, in the setup of the
quadrupolar geometry was employed to study the feasibility of
observing the dynamic phase transition for typical experimental
conditions. The frequency variation around the resonance leads to
the different behavior of the order parameter, depending on whether
the detuning is red- or blue- shifted. The obtained results can be
used for the experimental realization of the dynamic phase
transition in the process of resonant generation of topological
coherent modes.

Finally, we shall mention that several calculation using Bogoliubov
theory performed by Dziarmaga and Sacha \cite{Dziarmaga03} shown
that the effect of quantum and thermal depletion could be a serious
limitation on quantum coherence of atomic BEC. Second question is
the necessity of an efficient atomic population transference. If
both, quantum and thermal depletion, affect the process, we cannot
guarantee that the atoms which populates the excited state are a
Bose-Einstein condensate (a coherent state). Following Ruostekoski
and Walls \cite{Ruostekoski98}, the effects of decoherence due to
noncondensed atoms on BECs shows that purity decays fast. Hence, one
must have the interaction parameter much smaller than the
decoherence time scale associated with the nonlinear parameter.
Because our analysis shown that the macroscopic population of the
excited state is associated with small values of b, we believe the
effects of decoherence will be small.

\begin{acknowledgments}
This work was supported by Capes, Fapesp (Fundação de Amparo à
pesquisa do Estado de São Paulo), CNPq (Conselho Nacional de
Desenvolvimento Científico e Tecnológico) and Fapemig (Fundação de
Amparo à pesquisa de Minas Gerais). We are grateful to E.P. Yukalova
for many useful discussions.
\end{acknowledgments}


\end{document}